\documentclass[fleqn,12pt]{article}
\usepackage{epsfig}
\begin{document}
\textheight 22cm 
\textwidth 15cm 
\noindent 
{\Large \bf Zonal flow generation in collisionless trapped electron mode turbulence}
\newline
\newline
J. Anderson\footnote{anderson.johan@gmail.com}, H. Nordman, R. Singh*, J. Weiland
\newline
Department of Radio and Space Science, EURATOM-VR Association
\newline
Chalmers University of Technology, SE-41296 G\"{o}teborg, Sweden
\newline
*Institute for Plasma Research
\newline
Bhat, Gandhinagar
\newline
Gujarat, India 382428
\newline
\newline
\begin{abstract}
\noindent In the present work the generation of zonal flows in
collisionless trapped electron mode (TEM) turbulence is studied
analytically. A reduced model for TEM turbulence is utilized based
on an advanced fluid model for reactive drift waves. An analytical
expression for the zonal flow growth rate is derived and compared
with the linear TEM growth, and its scaling with plasma parameters
is examined for typical tokamak parameter values.
\end{abstract}
\newpage
\section{Introduction}

The study of the generation and suppression of turbulence and
transport in tokamak plasmas is still a high priority topic in
theoretical and experimental magnetic fusion research. In recent
studies, the important role played by nonlinearly self-generated
zonal flows for the regulation of turbulent transport has been
emphasized~\cite{a11}-~\cite{a13}. These are radially localized
flows (wave-vector $q=(q_x,0,0)$), propagating mainly in the
poloidal direction, which can reduce the radial transport by
shearing the eddies of the driving background turbulence.

The turbulence and anomalous transport observed in tokamak plasmas
is generally attributed to short-wavelength drift-type
instabilities, driven by gradients in the plasma density,
temperature, magnetic field, etc. For the hot core region of a
tokamak plasma, the two main drift-wave candidates are the
toroidal Ion Temperature Gradient (ITG) Mode and the collisionless
Trapped Electron Mode (TEM).

Accordingly, the generation of zonal flows by non-linear
interactions among drift waves has recently been studied both
analytically~\cite{a14}-~\cite{a22} and numerically
~\cite{a23}-~\cite{a35}. While a substantial effort has been devoted to
the study of zonal flow generation by ITG modes (see e.g. ~\cite{a16},
~\cite{a18} and~\cite{a20} for an analytical treatment), very
little work has been published on zonal flows driven by pure TEM
~\cite{a27}-~\cite{a29}.
 In tokamak plasma experiments, TEM are expected to play a dominant role
in the hot electron regime ($T_e>T_i$), relevant for experiments with dominant
 central electron heating ~\cite{a61}-~\cite{a62} and in advanced confinement 
regimes with electron transport
barriers ($\eta_e>\eta_i$). The study of zonal flows and turbulence driven by 
TEM is therefore crucial for the assessment of these regimes. In addition,
an estimate of the zonal flow generation close to marginal stability is 
essential in order to discriminate between
the TEM and the potentially important Electron Temperature Gradient
(ETG-) mode~\cite{a63}-~\cite{a64} in comparison of experimental profiles 
against linear thresholds. The experimental temperature gradients (or 
inverse scale lengths) are typically found to be about a factor of 2 above 
the linear thresholds, both for ITG and TEM. In the Cyclone study~\cite{a35},
this factor was 1.7 for the ITG mode. However, the nonlinear upshift, due to 
zonal flows,
increased the effective ITG threshold by a factor 1.5, thus bringing it much
closer to the experimental gradient. Although a comprehensive and
quantitative investigation of zonal flow generation would require a nonlinear
gyrokinetic treatment, a qualitative analytical study, based on a reduced
set of fluid equations, is feasible and also more transparent in terms of
physics interpretation.

In the present paper, the generation of zonal flows by pure TEM is
studied analytically in the limit $\eta_i = 0$ where the ITG mode is 
suppressed, using a reduced fluid model for the trapped electron dynamics. 
A system of equations is derived which describes the coupling between the 
 background TEM turbulence, described by a wave-kinetic equation, and the 
zonal flow modes generated by Reynolds stress forces. The qualitative 
analytical technique used here follows closely the WKB analysis employed in
~\cite{a18} and ~\cite{a20} for zonal flow generation by ITG
turbulence. The purpose of the study is to obtain a qualitative
estimate of the zonal flow growth rate driven by TEM and to
compare with an ITG driven case, and in addition examine its
scaling with plasma parameters for typical tokamak parameter
values.

The paper is organized as follows. In Section II the model
equations for the ITG/TEM system is presented and a reduced model
for TEM turbulence is presented. The equations describing the
coupling between the background TEM turbulence and the zonal flows
are presented in Section III and most explicit derivations are put in the 
Appendix A.
 In Section IV the results are discussed and finally there is a summary in 
Section V.

\section{Reduced Model for Trapped Electron Modes}
The description used for the coupled toroidal ITG and
collisionless TEM system is based on the continuity and
temperature equations for the ions and the trapped electrons~\cite{a36}:
\begin{eqnarray}
\frac{\partial n_{j}}{\partial t} +\nabla \cdot \left( n_{j} \vec{v}_{E} + 
n_{j} \vec{v}_{\star j} \right)+ \nabla \cdot  \left( n_j \vec{v}_{Pj} + 
n_j \vec{v}_{\pi j}\right)  & = & 0 \\
\frac{3}{2} n_{j} \frac{d T_{j}}{dt} + n_{j} T_{j} \nabla \cdot \vec{v}_{j} + 
\nabla \cdot \vec{q}_{j} & = & 0 \\
q_j =  \frac{5}{2} \frac{p_j}{m_j \Omega_{cj}} \left( e_{\parallel} 
\times \nabla T_j\right)
\end{eqnarray}
where $n_j$, $T_j$ are the density and temperature perturbations
($j=i$ and $j=et$ represents ions and trapped electrons) and 
$\vec{v}_j = \vec{v}_E + \vec{v}_{\star} + \vec{v}_{Pj} + \vec{v}_{\pi j}$, 
$\vec{v}_E$ is the $\vec{E} \times \vec{B}$ velocity,
$\vec{v}_{\star}$ is the diamagnetic drift velocity, $\vec{v}_{Pj}$ is the 
polarization drift velocity, $\vec{v}_{\pi j}$ is the stress tensor drift 
velocity and $\vec{q}_j$ is the heat flux. The derivative is defined
as $d/dt = \partial / \partial t + \rho_s c_s \vec{z} \times
\nabla \tilde{\phi} \cdot \nabla$ and $\phi$ is the electrostatic
potential. In the forthcoming equations $\tau = T_e/T_i$,
$\vec{v}_{\star} = \rho_s c_s \vec{y}/L_n $, $\rho_s = c_s/\Omega_{ce}$ where 
$c_s=\sqrt{T_e/m_i}$, $\Omega_{ce} = eB/m_e
c$. We also define $L_f = - \left( d ln f / dr\right)^{-1}$,
$\eta_j = L_n / L_{T_j}$, $\omega_{Dj}/\omega_{*j} = \epsilon_n
g_j = \frac{2 L_n}{R} g_j$, where $R$ is the major radius,
$\alpha_i = \frac{ \left( 1 + \eta_i\right)}{\tau}$ and $g_j$
represents the variation of $\omega_{Dj}$ along the field line. The 
geometrical quantities are calculated
in the strong ballooning limit ($\theta = 0 $, $g_j$ = 1).
The perturbed field variables are normalized as 
$\tilde{\phi} = (L_n/\rho_s) e \delta \phi /T_e$, $\tilde{n} = 
(L_n/\rho_s) \delta n /
n_0$, $\tilde{T}_{j} = (L_n/\rho_s) \delta T_{j} / T_{e0}$. 
The perpendicular length scale and time are normalized to $\rho_s$ and
$L_n/c_s$, respectively. Equations (1) and (2) can now be simplified to
\begin{eqnarray}
\frac{\partial \tilde{n}_{et}}{\partial t} + f_t \frac{\partial 
\tilde{\phi}}{\partial y} + \epsilon_n g_e \frac{\partial}{\partial y} 
\left(-f_t \tilde{\phi} + \tilde{n}_{et} + f_t \tilde{T}_{et} \right) 
& = & \nonumber \\ 
- \left[\phi,n_{et} \right] \\ 
\frac{\partial \tilde{T}_{et}}{\partial t} + \frac{5}{3} 
\epsilon_n g_e \frac{\partial \tilde{T}_{et}}{\partial y} + 
\left( \eta_e - \frac{2}{3}\right)\frac{\partial 
\tilde{\phi}}{\partial y} - \frac{2}{3 f_t} \frac{\partial 
\tilde{n}_{et}}{\partial t} & = & \nonumber \\
- \left[\phi,T_{et} \right] + \frac{2}{3f_t} \left[\phi,n_{et} \right] \\
\frac{\partial \tilde{n}_{i}}{\partial t} - \left(\frac{\partial}{\partial t} 
- \alpha_i \frac{\partial }{\partial y}\right) \nabla^2 \tilde{\phi} + 
\frac{\partial \tilde{\phi}}{\partial y} - \epsilon_n g_i 
\frac{\partial}{\partial y} \left( \tilde{\phi} + \frac{1}{\tau} 
\left(\tilde{n_{i}} + \tilde{T}_{i} \right) \right) & = &\nonumber \\
- \left[\tilde{\phi},\tilde{n}_i \right] + \left[\tilde{\phi}, 
\nabla^2_{\perp} \tilde{\phi} \right] + \frac{1}{\tau} \left[\tilde{\phi}, 
\nabla^2_{\perp} \left( \tilde{n}_i + \tilde{T}_i\right) \right] \\
\frac{\partial \tilde{T}_{i}}{\partial t} - \frac{5}{3 \tau} \epsilon_n g_i 
\frac{\partial \tilde{T}_{i}}{\partial y} + \left( \eta_i - 
\frac{2}{3}\right)\frac{\partial \tilde{\phi}}{\partial y} - 
\frac{2}{3} \frac{\partial \tilde{n_{i}}}{\partial t} & = & \nonumber \\
- \left[\tilde{\phi},\tilde{T}_{i} \right] + \frac{2}{3} 
\left[\tilde{\phi},\tilde{n}_i \right].
\end{eqnarray}
Here $f_t=n_{et}/n_0$ is the fraction of trapped electrons. 
The Poisson bracket is
$\left[ A ,B \right] = \partial A/\partial x \partial B/\partial y
- \partial A/\partial y \partial B/\partial x$. The system is
closed using the quasineutrality condition
\begin{eqnarray}
\delta n_i = \delta n_{e} & = & \delta n_{et} + \delta n_{ef}.
\end{eqnarray}
where a Boltzmann distribution has been assumed for the free
electrons. After linearizing equations (4)-(7), the dispersion
relation for the coupled ITG/TEM system is obtained as
\begin{eqnarray}
0 & = & \frac{\omega_{\star}}{N_i} \left[ \omega 
\left( 1 - \epsilon_n g_i\right) - \left( \frac{7}{3} - 
\eta_i - \frac{5}{3} \epsilon_n g_i\right) \omega_{Di} \right. \nonumber \\
& - & \left. k_y^2 \rho_s^2\left( \omega - \omega_{\star i} 
\left( 1 + \eta_i\right)\right)\left( \frac{\omega}{\omega_{\star e}} + 
\frac{5}{3 \tau} \epsilon_n g_i\right)\right] \nonumber \\
& - & f_t \frac{\omega_{\star e}}{N_e} \left[ \omega \left( 1 -
\epsilon_n g_e\right) - \left( \frac{7}{3} - \eta_e - \frac{5}{3}
\epsilon_n g_e\right)\omega_{De}\right] - 1 + f_t
\end{eqnarray}
where
\begin{eqnarray}
N_j = \omega^2 - \frac{10}{3} \omega \omega_{D j} + \frac{5}{3} \omega_{D j}^2
\end{eqnarray}
Depending on the plasma parameters, the dispersion relation (9)
may contain 0, 1 or 2 unstable modes. For modes propagating in the
ion drift direction (usually the ITG mode), $N_i < N_e$, while for
modes propagating in the electron drift direction (TEM), $N_e
< N_i$. The modes become de-coupled when the inequalities are
strong. Thus, for $N_e >> N_i$ we obtain a pure ITG mode. 
For pure ITG mode physics, the fluid model used here has been found to be in
good qualitative agreement with a number of gyrokinetic treatments. For
example, both the $\eta_i$-scaling of the ion heat transport~\cite{a35} and 
the nonlinear upshift of the linear ITG threshold due to zonal
flows~\cite{a35} has been recovered by the fluid model~\cite{a38}. 
A more comprehensive version of the model, based on the
full ITG and TE system (Equations 4-7), has been heavily used in predictive
transport code simulations~\cite{a39}-~\cite{a41} of tokamak discharges. 
The simulation results indicate that the model is able to reproduce 
experimental profiles of temperatures and density, inside the edge region, 
with good accuracy over a wide range of plasma parameters. In the limit 
$N_i >> N_e$, we obtain a pure TEM. In tokamak plasmas the TEM is expected 
to dominate in the hot electron regime ($T_e>>T_i$) and/or in regimes with 
$\eta_e>>\eta_i$. However, for peaked density profiles (small $\epsilon_n$), 
the ion and trapped electron responses are strongly coupled 
($N_i \approx N_e$) and a density gradient driven TEM appears for weak 
temperature gradients. In the following we will neglect the effects of ion 
perturbations on the TEM and hence only consider electron temperature 
gradient driven TEM (not including the electron temperature gradient mode 
(ETG)). The dispersion relation then takes the form

\begin{eqnarray}
0 & = & \omega^2 + \omega k_y \left( \xi \left( 1-\epsilon_n g \right) - 
\frac{10}{3}\epsilon_n g\right) \nonumber \\
& + &  k_y^2 \epsilon_n g \left(\xi \left(\eta_e -\frac{2}{3} \epsilon_n g - 
\frac{7}{3} \left(1 - \epsilon_n g \right)\right)+ \frac{5}{3} \epsilon_n g
\right) \\
\xi & = & \frac{f_t}{1 - f_t}
\end{eqnarray}
Equation 11 describes TEMs driven by $R/L_{Te}$ and suppressed by $R/L_n$ 
leading to a linear TEM stability threshold in the parameter 
$\eta_e = L_n/L_{Te}$. In the considered limit the TEM is fairly symmetrical 
to the toroidal ITG mode, except that effects of finite-Larmor-radius and 
parallel electron dynamics do not appear in the TEM dispersion relation. 
The solution to Equation 11 is given by
\begin{eqnarray}
\omega_r & = & - \frac{k_y}{2} \left( \xi \left(1 - \epsilon_n g\right) - 
\frac{10}{3} \epsilon_n g \right)  \\
\gamma & = & k_y \sqrt{ \xi \epsilon_n g \left( \eta_e -\eta_{eth}\right)}
\end{eqnarray}
where $\omega = \omega_r + i \gamma$ and the linear stability threshold is 
given by
\begin{eqnarray}
\eta_{e th} = \frac{2}{3} - \frac{\xi }{2} + \frac{10}{9 \xi }\epsilon_n g
+ \frac{\xi \epsilon_n g}{4} + \frac{\xi }{4 \epsilon_n g}.
\end{eqnarray}
In this regime we can define a reduced model for electron temperature 
gradient driven TEM turbulence by
retaining equations (4)-(5) for the trapped electron fluid while
neglecting ion dynamics (6)-(7) (see Appendix A). The
effect of the neglected ion dynamics on the linear physics will in
the following be quantified by comparing the results of the
reduced (equation (11)) with the complete (equation (9)) dispersion relation.

\section{Zonal Flow Generation}
In describing the large scale plasma flow dynamics it is assumed
that there is a sufficient spectral gap between the small scale
TEM fluctuations and the large scale flow. The electrostatic
potential is represented by
\begin{eqnarray}
\phi(X,x,y,T,t) = \Phi(X,T) + \tilde{\phi}(x,y,t)
\end{eqnarray}
where $\tilde{\phi}(x,y,t)$ is the fluctuating potential varying on the 
turbulent scales $x,y,t$ and $\Phi(X,T)$ is the zonal flow potential varying 
on the slow scale $X,T$ (the zonal flow potential is independent on $Y$). 

The evolution of the TEM turbulence in the background of the slowly
varying zonal flow $\Phi(X,T)$ can be described by the
wave-kinetic equation~\cite{a16},~\cite{a51} and~\cite{a52}
for the adiabatic invariant $N_k = C_k |\tilde{\phi}_k|^2$ (see
Appendix A for a derivation of $N_k$ and $C_k$).
\begin{eqnarray}
\frac{\partial }{\partial t} N_k(x,y,t) & + & \frac{\partial }{\partial k_x} 
\left( \omega_k + \vec{k} \cdot \vec{v}_0 \right)\frac{\partial 
N_k(x,y,t)}{\partial x} - \frac{\partial }{\partial x} \left( \vec{k} 
\cdot\vec{v}_0\right) \frac{\partial N_k(x,y,t)}{\partial k_x} \nonumber \\
& = &  \gamma_k N_k(x,y,t) - \Delta\omega N_k(x,y,t)^2
\end{eqnarray}
Here, $\vec{v}_0$ is the zonal flow part of the ExB drift. In this analysis 
it is assumed that the RHS is approximately zero (stationary turbulence). 
The role of non-linear interactions among the TEM fluctuations (here 
represented by a non-linear frequency shift $\Delta\omega$) is to balance 
linear growth rate, i.e. $ \gamma_k N_k(x,y,t) - \Delta\omega N_k(x,y,t)^2 
\approx 0$. The TEM turbulence is assumed to be adiabatically modulated by 
the slowly growing potential $\Phi(X,T)$. Equation (17) is then expanded under
the assumption of small deviations from the equilibrium spectrum
function; $N_k = N_k^0 + \tilde{N}_k$ where $\tilde{N}_k$ evolves
at the zonal flow time and space scale  $\left( \Omega, q_x, q_y = 0\right)$, 
as
\begin{eqnarray}
- i \left(\Omega - q_x v_{gx} + i \gamma_k\right) \tilde{N}_k = k_y 
\frac{\partial^2 }{\partial x^2} \Phi \frac{\partial N_k^0
}{\partial k_x}
\end{eqnarray}
\begin{eqnarray}
\tilde{N}_k = - q_x^2 k_y \frac{\partial N_k^0}{\partial k_x}
\frac{i}{\Omega - q_x v_{gx} + i \gamma_k}\Phi
\end{eqnarray}
Here $v_{gx} = \partial \omega/\partial k_x \approx 0$, since the effects of 
electron FLR is neglected.
\\
\vspace{0.5cm}
\\
The evolution equations for the zonal flow is obtained after
averaging the ion-continuity equation over the magnetic flux
surface and over fast scales and employing quasi-neutrality (equation 8)
\begin{eqnarray}
\frac{\partial}{\partial t} \nabla_x^2 \Phi  = \nabla_x^2 
\left<\frac{\partial}{\partial
x} \tilde{\phi}_k \frac{\partial}{\partial y}\tilde{\phi}_k
\right>
\end{eqnarray}
Here we have assumed that the turbulence is dominated by the TEM 
($\tilde{n_i} << \tilde{n}_{et} $) and hence only the
small scale self interactions among the TEM are contributing
to the Reynolds stress in the RHS of (20) ~\cite{a53}. Expressing
the Reynolds stress terms in equation (20) in $N_k$ we obtain
\begin{eqnarray}
-i \Omega \Phi = \int d^2 k k_x k_y C_k^{-1} N_k
\end{eqnarray}
The factor $C_k$ defines the relationship between small scale turbulence 
and the wave action density, see Appendix equation A9 for details.
Integrating by parts in $k_x$ and assuming a monochromatic wave
packet $N_k^0 = N_0 \delta\left(k - k_0\right)$ and using
equations (19) and (21) gives
\begin{eqnarray}
\Omega^2 = - q_x^2 C_k^{-1} k_y^2 N_0
\end{eqnarray}
The dispersion relation for zonal flow $\Omega$ reduces to
\begin{eqnarray}
\Omega = i q_x k_y\sqrt{C_k^{-1} N_0}.
\end{eqnarray}
Hence, the zonal flow growth rate scales as $\Omega \propto |\phi|_k$. 
In expressing the zonal flow growth in dimensional form making use
of equation (23), it is assumed that the background turbulence (in
the absence of zonal flows) reach the mixing length level for temperature 
gradient driven modes corresponding to $\tilde{T}_e = \frac{1}{k_x L_{T_e}}$. 
We then obtain (see Appendix A)
\begin{eqnarray}
\Omega  & = & i q_x k_y F \frac{\eta_e}{k_x L_n} \\
F & = & \frac{\sqrt(\Delta_k^2 + \gamma_k^2)}{|\eta_e - \frac{2}{3 \xi} 
\epsilon_n g \left( 1 + \xi\right)|}
\end{eqnarray}
where $q_x$ is the zonal flow wave number, $k_y$ is the TEM wave number, 
$\Delta_k = - \frac{k_y}{2}\left(\xi - \epsilon_n g \xi + \frac{4}{3}
\epsilon_n g\right)$ and $\gamma_k$ is the linear TEM growth rate. 
The function $F$ is usually large in regions close to marginal stability 
due to the denominator in equation (25). This is a result of the 
quasilinear treatment of the TEM perturbations $\phi_k$ (appearing in 
equation 23) and $\tilde{T}_{e} = 1/(k_x L_{Te})$. 
\section{Results and discussion}
An algebraic equation (24) describing the zonal flow growth rate
driven by short-wavelength TEM turbulence has been derived.
The zonal flow growth rates will in the following be calculated and 
compared to the linear TEM growth rates. First, the linear TEM descriptions 
are compared. In Figure 1, the solutions to the full system of 4
equations describing the coupled ITG/TE system (squares, equation (9)) is 
compared to the reduced model with 2 equations (asterisks, equation (11)) 
for the pure TEM. The results are shown for $\eta_i$ =0, $\tau = 1$,
$\epsilon_n =1.0$, $k \rho = 0.3$ and $f_t = 0.5$. 
The $\eta_e$ scalings for the TEM
eigenvalues are found to be in good qualitative agreement in this regime 
(for fairly flat density profiles); the growth rates are within $20\%$ 
except close to the linear threshold. The influence of the ion dynamics 
on the TEM stability would be reduced further in the limit of cold ions 
($\tau>>1$).

\begin{figure}
  \includegraphics[height=.3\textheight]{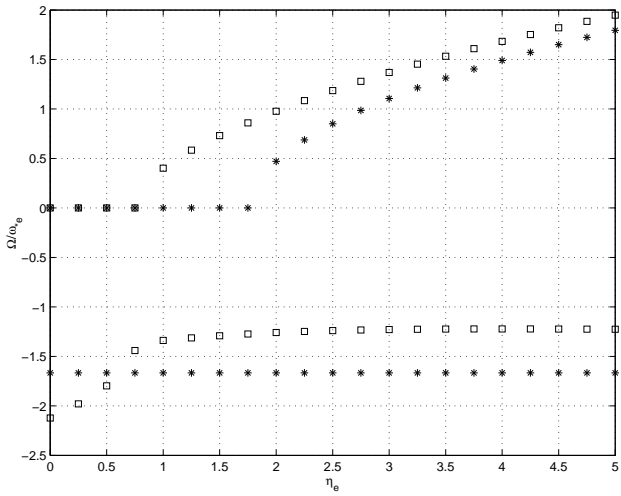}
  \caption{The solutions for the TEM eigenvalues using the full system
of 4 equations (squares) is compared to the reduced model with 2 equations 
(asterisks). The growth rate and the real frequency with reversed sign
(normalized the electron diamagnetic drift frequency) vs $\eta_e$
is displayed. The results are shown for $\tau = 1$, $\epsilon_n = 1.0$, 
$\eta_i$ = 0 and $k\rho = 0.3$.}
\end{figure}

Next, the zonal flow growth rate is studied. In Figure 2 the effects of 
$\eta_e$ and $\epsilon_n$ on zonal flow growth rate (normalized to 
the TEM growth rate) are displayed. The other parameters are $f_t = 0.5$, 
$\eta_i = 0$, $\tau = 1$ and $k_x \rho = k_y \rho = q_x \rho = 0.3$. 
The results for the zonal flow growth rate are shown for $\epsilon_n = 0.5$ 
(with $\eta_{eth} = 1.38$, boxes), $\epsilon_n = 0.7$ 
(with $\eta_{eth} = 1.48$, rings). There is a significant increment in the 
zonal flow growth rate (normalized to the linear TEM growth rate) just 
above marginal stability. Part of this increment is due to the reduction of 
the linear TEM growth. In addition, a resonance in the analytical expression 
for the ZF growth rate (see equation 25) is obtained just above the linear 
TEM threshold  which further enhances the ZF growth in this region. A similar 
resonance was found for zonal flows driven by ITG modes ~\cite{a20} and 
~\cite{a21}.

\begin{figure}
  \includegraphics[height=.3\textheight]{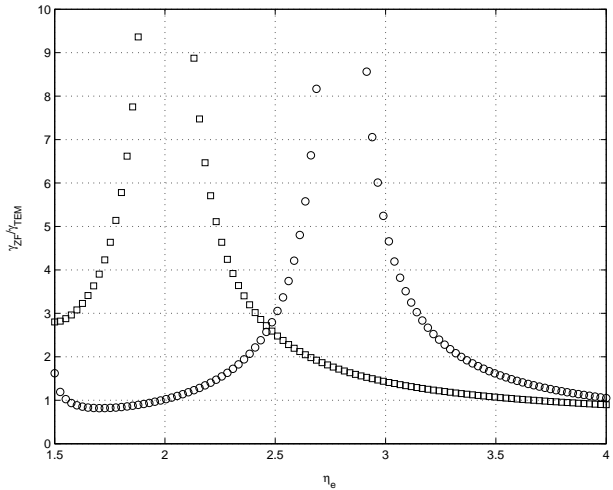}
  \caption{The zonal flow growth rate (normalized to the linear TEM
growth rate) vs $\eta_e$ with $\epsilon_n$ as parameter is
displayed. The results are shown for $\epsilon_n = 0.5$ (squares), 
$\epsilon_n = 0.7$ (rings). The other parameters are $\eta_i = 0$, 
$\tau = 1$ and $k_x \rho = k_y \rho = q_x \rho = 0.3$ and $f_t = 0.5$.}
\end{figure}

Next the zonal flow growth rates (normalized to the TEM growth rate) as a 
function of $\epsilon_n$ ($=2L_n/L_B$) are displayed (derived from equation. 
24) with $f_t$ (the fraction of trapped electrons) as a parameter. 
In Figure 3, the results are shown for $f_t = 0.5$ (asterisks) and 
$f_t = 0.7$ (boxes). The other parameters are as in Figure 2 with 
$\eta_e = 3$. Similar to the results in Figure 2 a strong growth of 
zonal flow is obtained close to the $\epsilon_n$ threshold (for 
($f_t = 0.5$, $\epsilon_{n th} \approx 0.09$) and ($f_t = 0.7$, 
$\epsilon_{n th} \approx 0.18$)). We emphasize the the reduced TEM model 
used here does not include the $\nabla n$ drive which will be important 
in the region of small $\epsilon_n$ ($\epsilon_n << 1$).

\begin{figure}
  \includegraphics[height=.3\textheight]{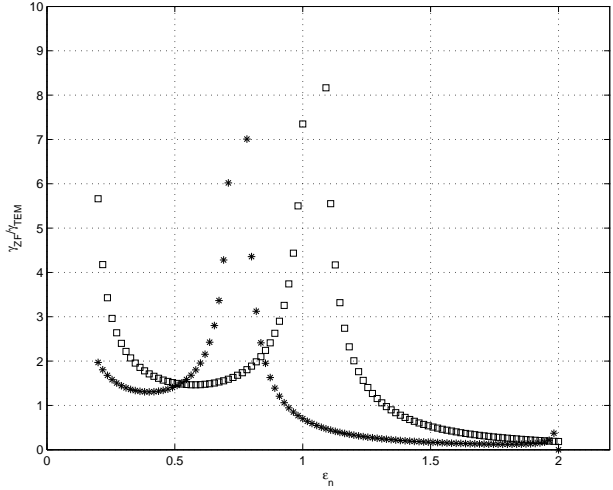}
  \caption{The zonal flow growth rate (normalized to the linear TEM growth 
rate) vs $\epsilon_n$ ($=2L_n/L_B$) with fraction of trapped electrons 
($f_t$) as parameter. The other parameters are as in Figure 2 with 
$\eta_e = 3$,  The results are shown for $f_t = 0.5$ (asterisks) and 
$f_t = 0.7$ (squares).}
\end{figure}

Figure 4 compares the zonal flow growth rate generated by pure TEM turbulence 
for $\eta_e = 3$, $\eta_i = 0$ to that generated by ITG mode turbulence for 
$\eta_e = 0$, $\eta_i = 3$, see equation (33) in Ref.~\cite{a20} (note the 
difference in the definition of $\tau$) as a function of $\tau$ with 
$\epsilon_n$ as parameter. The comparison is done assuming equal saturation 
levels for the ITG mode and TEM cases. The results are shown for 
$\epsilon_n = 0.5$ (diamonds), $\epsilon_n = 1.0$ (plus) and 
$\epsilon_n = 1.5$ (squares).
In the case of cold ions (small $1/\tau$), the ITG and TEM
turbulence generates comparable levels of ZF growth, whereas for
equal electron and ion temperatures, the ITG mode generates
significantly larger levels of zonal flows (typically around 2
times larger ZF growth rates are obtained for the ITG case). This
is due to the nonlinear diamagnetic effects which significantly
contribute to the ITG driven ZF growth~\cite{a16}. The relatively 
weak generation of zonal flows obtained here by TEM is consistent 
with recent results from nonlinear gyrokinetic simulations of TEM 
turbulence~\cite{a29}. 

\begin{figure}
  \includegraphics[height=.3\textheight]{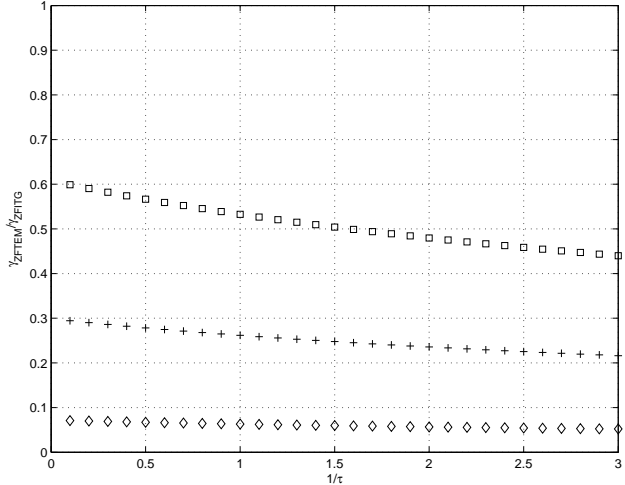}
  \caption{The zonal growth rate generated from TEM turbulence
(for $\eta_e = 3$, $\eta_i = 0$) compared to that generated from
ITG mode turbulence (for $\eta_i = 3$, $\eta_e = 0$) as a function
of $\tau$ with $\epsilon_n$ as parameter. The results are shown
for ($\epsilon_n = 0.5$) (diamonds),
($\epsilon_n = 1.0$) (plus) and ($\epsilon_n = 1.5$)
(squares).}
\end{figure}

\section{Summary}
The present paper investigates the generation of zonal flows by 
collisionless trapped electron modes (TEM). An algebraic equation 
which describes the zonal flow growth rate in the presence of collisionless 
TEM turbulence is derived and solved numerically in the strong
ballooning limit. A reduced model for the electron temperature gradient 
driven TEM is utilized based on an advanced fluid model including the 
trapped electron continuity and the electron temperature equations 
while neglecting the influence of ion perturbations. The generation of 
zonal flows is described by the vorticity equation and the time evolution 
of the TEM turbulence in the presence of the slowly growing zonal flow is
described by a wave kinetic equation.

It is found that the reduced TE model (2-equations) is
qualitatively able to reproduce the linear physics of the full model
(4-equations) in the TEM dominated regimes where $T_e>>T_i$ and/or in
regimes where $\eta_e>>\eta_e$. For reasonable flat density profiles, 
the linear TE growth rates are typically
within $20\%$ except close to the linear stability threshold.

There is a significant increment in the zonal flow growth rate
(normalized to the linear TEM growth) close to the linear TEM
threshold where a resonance in the 
ZF generation is obtained. This may result in a larger level of zonal flow,
 and consequently a lower level of TEM turbulence, in this region.

A qualitative comparison of the zonal flow growth rate generated
by pure TE and ITG mode turbulence shows that the ITG mode
generates significantly larger levels of ZF growth except in the
case of cold ions (small $1/\tau$) where the TEM driven ZF growth
is comparable to the ITG case. 

In the present paper the generation of zonal flows has been studied
 analytically by analysing the linear zonal flow growth rates. A complete 
analysis should also involve an assessment of the stability properties of 
the generated flows. This is more suitable for numerical analysis and is 
outside the scope of the present paper. In addition, a more comprehensive 
comparison of TEM and ITG driven zonal flows should include cases where 
$\eta_e \approx \eta_i$ and/or cases where the modes are strongly coupled 
(for peaked density profiles), using the full ITG/TEM system. This is left 
for future work.

\appendix
\thispagestyle{empty}
\renewcommand\theequation{\thesection.\arabic{equation}}
\setcounter{equation}{0}
\section{Adiabatic Invariant in TEM turbulence}
In this Appendix the derivation of the adiabatic invariant in TEM
driven turbulence is presented. The method has been described in
detail in Ref.~\cite{a16} and ~\cite{a20} (and References therein)
and only a brief summary is given here. From equations 4 and 5 we
get,
\begin{eqnarray}
\frac{\partial \tilde{n}_{et}}{\partial t} - \xi \frac{\partial
\tilde{n}_{et}}{\partial y } + \epsilon_n g \frac{\partial
}{\partial y} \left(\xi \tilde{n}_{et} + \tilde{n}_{et} + f_t
T_{et}\right) & = & \nonumber \\
 - \left[\phi, \tilde{n}_{et}\right] \\
\frac{\partial \tilde{T}_{et} }{\partial t} + \frac{7}{3}
\epsilon_n g \frac{\partial \tilde{T}_{et}}{\partial y} -
\left(\eta_e - \frac{2}{3} \epsilon_n g \right)
\frac{\xi}{f_t} \frac{\partial \tilde{n}_{et}}{\partial y} +
\frac{2}{3 f_t} \epsilon_n g \frac{\partial
\tilde{n}_{et}}{\partial y} & = & \nonumber \\
- \left[\phi,\tilde{T}_{et}\right].
\end{eqnarray}
Here, the interaction between the TEM perturbations have been
omitted (see discussion after equation 16 for this.) It should be noted 
that the relationship between the electrostatic potential $\tilde{\phi}$ 
and the trapped electron density $\tilde{n}_{et}$ is given by 
$\tilde{\phi} = -\frac{1}{1-f_t} \tilde{n}_{et}$ (see equation 8).
To determine the generalized wave action density $N_k = |\Psi_k|^2$ 
we introduce
the normal coordinates  $\Psi_k = \tilde{n}_{et k} + \alpha_k
\tilde{T}_{et}$, where $\alpha_k$ is to be calculated. Multiplying
equation A.2 by $\alpha_k$ and adding it to equation A.1 gives
\begin{eqnarray}
\frac{\partial }{\partial t} \left(\tilde{n}_{et k} + \alpha_k
\tilde{T}_{etk} \right) & + &
 \left(-\xi + \left(1 + \xi \right) \epsilon_n g -
 \alpha_k \frac{\xi}{f_t} \left( \eta_e - \frac{2}{3} \epsilon_n g \right)
 + \frac{2}{3f_t}\epsilon_n g \alpha_k \right)
 \frac{\partial \tilde{n}_{et k}}{\partial y} \nonumber \\
& + & \left( \frac{7}{3} \epsilon_n g \alpha_k + f_t \epsilon_n g \right)
 \frac{\partial\tilde{T}_{etk}}{\partial y} = -
\left[\Phi,\tilde{n}_{et k} + \alpha_k \tilde{T}_{etk} \right]
\end{eqnarray}
The normal coordinates are found if the equation is rewritten as
in Ref. ~\cite{a20}
\begin{eqnarray}
\frac{\partial \Psi_k }{\partial t} + V_k \frac{\partial \Psi_k
}{\partial y} = - \left[\Phi, \Psi_k \right]
\end{eqnarray}
where
\begin{eqnarray}
V_k & = & -\xi + \left(1 + \xi \right) \epsilon_n g -
 \alpha_k \frac{\xi}{f_t} \left( \eta_e - \frac{2}{3} \epsilon_n g \right) +
 \frac{2}{3f_t}\epsilon_n g \alpha_k \\
\alpha_k & = &  \frac{\frac{7}{3}\epsilon_n g \alpha_k + 
f_t \epsilon_n g }{V_k}
\end{eqnarray}
which gives
\begin{eqnarray}
\alpha_k = \frac{- \frac{1}{2} \left( \xi - \epsilon_n g \xi +
\frac{4}{3} \epsilon_n g\right) + i \sqrt{\xi \epsilon_n g
\left(\eta_e - \eta_{e th}\right)}}{\frac{\xi}{f_t} \left(\eta_e -
\frac{2}{3 \xi} \left(1 + \xi \right)\epsilon_n g \right)}
\end{eqnarray}
The linear relations between $\tilde{\phi}_k$, $\tilde{n}_{etk}$ and
\begin{eqnarray}
\tilde{T}_{ek} = \frac{k_y \left(\eta_e - \frac{2}{3 \xi} \epsilon_n g_e 
(1+\xi)\right)}{\omega - \frac{7}{3}\epsilon_n g_e k_y} \tilde{\phi}_k
\end{eqnarray}
enables one to express $\Psi_k$ and $N_k$ as
\begin{eqnarray}
\Psi_k & = & \tilde{n}_{et k} + \alpha_k \tilde{T}_{ek} = \frac{2 i 
\gamma_k - f_t \left( -\Delta_k + i \gamma_k\right)}{\Delta_k + i\gamma_k } 
\tilde{\phi}_k \\
N_k & = & |\Psi_k|^2 = \left( \frac{4 \left( 1 - 
f_t \right)\gamma_k^2}{\Delta_k^2 + \gamma_k^2} + f_t^2\right)|
\tilde{\phi}_k|^2  = C_k |\tilde{\phi}_k|^2\\
C_k & = &  \left( \frac{4 \left( 1 - f_t \right)\gamma_k^2}{\Delta_k^2 + 
\gamma_k^2} + f_t^2\right)\\
\Delta_k & = & - \frac{k_y}{2}\left(\xi - \epsilon_n g \xi + 
\frac{4}{3}\epsilon_n g\right) \\
\gamma_k & = & k_y \sqrt{\xi \epsilon_n g \left(\eta_e - \eta_{e th}\right)}
\end{eqnarray}
The equations A.8 - A.12 describe the normal variables $\Psi_k$, the
adiabatic invariant $N_k$ and the linear TEM growth rate,
respectively.


\newpage
\begin{thebibliography}{200}
\bibitem{a11} Hasegawa A, Mcclennan CG and Kodama Y 1979 Phys. Fluids {\bf 22} 2122
\bibitem{a12} Bell RE, Levinton FM, Batha SH 1998 {\it et al.} Phys. Rev. Lett. 
{\bf 81} 1429
\bibitem{a13} Biglari H, Diamond PH and Terry PW 1990 Phys. Fluids B {\bf 2} 1 
\bibitem{a14} Diamond PH and Kim YB 1991 Phys. Fluids B {\bf 5} 2343
\bibitem{a16} Smolyakov AI, Diamond PH, Medvedev MV 2000 Phys. Plasmas {\bf 7} 3987
\bibitem{a17} Chen L, Lin Z and White RB 2000 Phys. Plasmas {\bf 7} 3129
\bibitem{a18} Smolyakov AI, Diamond PH and Malkov M 2000 Phys. Rev. Lett. {\bf 84} 491
\bibitem{a19} Melkov MA, Diamond PH and Smolyakov AI 2001 Phys. Plasmas {\bf 8} 1553
\bibitem{a20} Anderson J, Nordman H, Singh R and Weiland J 2002
Phys. Plasmas {\bf 9} 4500
\bibitem{a21} Mahajan S, Weiland J 2000 Plasma Phys. Contr. Fusion {\bf 42} 987
\bibitem{a22} Krommes JA, Kim CB 2000 Phys. Rev. E {\bf 62} 8508 
\bibitem{a23} Lin Z, Hahm TS, Lee WW, Tang WM and White RB 1998 Science 281 1835
\bibitem{a24} Hasegawa A and Wakatani M 1987 Phys. Rev. Lett. {\bf 59} 1581
\bibitem{a25} Nordman H, Weiland J, Jarmen A 1990 Nucl. Fusion {\bf 30} 983
\bibitem{a26} Hahm TS et al 1999 Phys. Plasmas {\bf 6} 922
\bibitem{a27} Ernst DR et al 2004 Phys. Plasmas {\bf 11} 2637
\bibitem{a29} Dannert T and Jenko F 2005 Phys Plasmas {\bf 12} 072309
\bibitem{a30} Lin Z, Hahm TS, Lee WW, Tang WM and Diamond PH 1999 Phys. Rev. Lett. 
{\bf 83} 3645
\bibitem{a31} Dimits A, Williams TJ, Byers JA and Cohen BI 1996 Phys. Rev. Lett. 
{\bf 77} 71
\bibitem{a32} Hammet G, Beer MA, Dorland W, Cowley SC and Smith SA 1993 Plasma 
Phys. Controlled Fusion {\bf 35} 937 
\bibitem{a33} Waltz RE, Kerbel GD and Milovich AJ 1994 Phys. Plasmas {\bf 1} 2229
\bibitem{a34} Beer MA 1995 Ph.D. dissertation, Princeton Univ.
\bibitem{a35} Dimits A, Bateman G, Beer MA et al. 2000 Phys. Plasmas {\bf 7} 969
\bibitem{a61} Ryter et al 2003 Nucl Fusion {\bf 43} 1396
\bibitem{a62} Angioni C et al 2005 Phys. Plasmas {\bf 12} 040701
\bibitem{a63} F. Jenko et al Phys Plasmas 7, 1904 2000
\bibitem{a64} R. Singh, P. Kaw and J. Weiland 2002 Nucl. Fusion {\bf 41} 1219 
\bibitem{a36} Weiland J 2000 Collective Modes in Inhomogeneous Plasmas, Kinetic and 
Advanced Fluid Theory (IOP Publishing Bristol) 115
\bibitem{a38} Dastgeer S, Mahajan S, Weiland J Phys. Plasmas 2002 {\bf 9} 4911 
\bibitem{a39} Weiland J and Nordman H 1991 Nucl. Fusion {\bf 31} 390
\bibitem{a40} Fr\"{o}jdh M, Strand P, Weiland J et al 1996 Plasma Phys Contr. 
Fusion {\bf 38} 325
\bibitem{a41} Strand P, Nordman H, Weiland J et al 1998 Nucl. Fusion {\bf 41} 545
\bibitem{a51} Vedenov AA, Gordeev AV, Rudakov LI 1967 Plasma Phys. {\bf 9} 719
\bibitem{a52} Smolyakov AI, Diamond PH 1999 Phys. Plasmas {\bf 6} 4410
\bibitem{a53} Horton W, Choi D, Terry P 1980 Phys. Fluids, {\bf 23} 590
\end{thebibliography}
\end{document}